\documentclass[prl,reprint,groupedaddress,showpacs,floatfix]{revtex4-1}
\usepackage[utf8]{inputenc}
\usepackage[pdftex]{graphicx} 
\usepackage{bm} 
\usepackage{amssymb} 
\usepackage{amsmath} 
\usepackage{hyperref}
\usepackage{csquotes}
\usepackage{etoolbox}
\apptocmd{\sloppy}{\hbadness 10000\relax}{}{}
\begin{document}
\title{Exact exchange-correlation kernels for optical spectra of model systems}
\date{\today}
\author{M.\ T.\ Entwistle}
\author{R.\ W.\ Godby}
\affiliation{Department of Physics, University of York, and European Theoretical Spectroscopy Facility, Heslington, York YO10 5DD, United Kingdom}

\begin{abstract}   
For two prototype systems, we calculate the exact exchange-correlation kernels $f_{\mathrm{xc}}(x,x',\omega)$ of time-dependent density functional theory. $f_{\mathrm{xc}}$, the key quantity for optical absorption spectra of electronic systems, is normally subject to uncontrolled approximation. We find that, up to the first excitation energy, the exact $f_{\mathrm{xc}}$ has weak frequency-dependence and a simple, though non-local, spatial form. For higher excitations, the spatial behavior  and frequency-dependence become more complex. The accuracy of the underlying exchange-correlation potential is of crucial importance.
\end{abstract}

\maketitle
Time-dependent Kohn-Sham density functional theory \cite{TDDFT, TDDFT3} (TDDFT) is in principle an exact and efficient theory of the excited-state properties of many-electron systems, including a wide variety of important spectroscopies such as optical absorption spectra of molecules and solids. However its application is restricted by the limitations of the available approximate functionals for electron exchange and correlation -- in particular the exchange-correlation kernel, $f_{\mathrm{xc}}$, the functional derivative of the exchange-correlation potential with respect to the electron density. To assist the construction of more powerful approximations for $f_{\mathrm{xc}}$, we calculate the \textit{exact} $f_{\mathrm{xc}}$ for small prototype systems, and analyze its character, including key aspects in which it differs from the common approximations.

In the Runge-Gross formulation \cite{TDDFT} of TDDFT the real system of interacting electrons is mapped onto an auxiliary system of noninteracting electrons moving in an effective local Kohn-Sham (KS) potential $v_{\mathrm{KS}} = v_{\mathrm{ext}} + v_{\mathrm{H}} + v_{\mathrm{xc}}$, with both systems having the same electron density $n$ at all points in space and time. Many TDDFT calculations are done within the framework of linear response theory, which describes how a system responds upon application of a weak, time-dependent external perturbation. The induced density is described by the interacting density-response function, the functional derivative $\chi = \delta n/ \delta v_{\mathrm{ext}}$. $\chi$ is related to the  non-interacting density-response function of the KS system, $\chi_{0} = \delta n/ \delta v_{\mathrm{KS}}$, by the Dyson equation \footnote{Matrix multiplication for the spatially non-local quantities $\chi$, $\chi_{0}$ and $f_{\mathrm{xc}}$, and $\omega$-dependence, are implied.} \cite{Dyson} $\chi = \chi_{0} + \chi_{0} (u+f_{\mathrm{xc}}) \chi$, where $u$ is the bare Coulomb interaction. $\chi_{0}$ is to be obtained from a ground-state DFT calculation. $\chi$ can then be used to compute, for example, the optical absorption spectrum of the system,
\begin{equation} \label{oas}
\sigma(\omega) = -\frac{4\pi \omega}{c}\int \int \text{Im} \big(\chi(x,x',\omega)\big) \ x \ x' \ dx \ dx'.
\end{equation}

In practice, both $v_{\mathrm{xc}}$ and its functional derivative $f_{\mathrm{xc}}$ must be approximated. While there have been some successes, the commonly used adiabatic TDDFT functionals, such as the adiabatic local density approximation \cite{ALDA_LR, ALDA_LR2} (ALDA), often fail in extended systems. For example, the optical absorption spectra of many semiconductors and insulators are not even qualitatively described, with excitonic effects and many-electron excitations omitted \cite{fxc_alda, fxc_alda2}, and the optical gap underestimated. Here, approximations for $f_{\mathrm{xc}}$ achieve little improvement over the random phase approximation (RPA), in which $f_{\mathrm{xc}}$ is neglected entirely \cite{fxc_rpa_lda}. Attempts to improve approximations for $f_{\mathrm{xc}}$ include exact-exchange methods \cite{fxc_exx, fxc_exx2, fxc_exx3, fxc_exx4, fxc_exx5, fxc_exx6}, diagrammatic expansions using perturbative methods \cite{fxc_pert, fxc_pert2}, and adding long-range contributions \cite{fxc_nonlocal, fxc_nonlocal2, fxc_nonlocal3, fxc_nonlocal4}. Another approach involves calculations of the homogeneous electron gas (HEG) \cite{fxc_heg, fxc_heg2, fxc_heg3, fxc_heg4, fxc_heg5}. Kernels derived from the Bethe-Salpeter equation \cite{fxc_bse, fxc_bse2, fxc_bse3, fxc_bse4, fxc_bse5} have had some success, but require a relatively expensive many-body perturbation theory calculation as their input, and are outside the KS TDDFT framework. 

There have been a limited number of studies conducted on analyzing the character of the exact $f_{\mathrm{xc}}$, all of which focus on its frequency-dependence. One approach has been to calculate the exact adiabatic $f_{\mathrm{xc}}$ for model systems \cite{fxc_sumrule2}, in order to investigate its performance upon application and deduce when memory effects become important. This approach has been used in simple Hubbard systems \cite{fxc_hubbard, fxc_hubbard2} and extended by analyzing additional properties, such as the frequency-dependence of the full $f_{\mathrm{xc}}$ around double excitations. Other research has explored how this frequency-dependence of $f_{\mathrm{xc}}$ turns the single-particle quantities of exact KS TDDFT into many-body excitations \cite{fxc_anharmonic} and its behaviour for long-range excitations has been analyzed in order to develop approximate kernels \cite{fxc_longrange}.

In this paper, we explore the properties of exact xc kernels, including full spatial and frequency-dependence, in order to inform the development of improved approximate functionals. We employ our iDEA code \cite{iDEA} which solves the many-electron Schr{\"o}dinger equation exactly for small, one-dimensional prototype systems \footnote{We perform calculations for systems of two spinless electrons interacting via the appropriately softened Coulomb repulsion \cite{SoftenedCoulomb} $u(x,x') = (|x-x'|+1)^{-1}$, and work in Hartree atomic units: $m_{\mathrm{e}} = \hbar = e = 4\pi\varepsilon_{0} = 1$.} \footnote{See Supplemental Material at `Link' for the parameters of the model systems, and details on our calculations to obtain converged results.}. From the many-electron eigenstates of the system we calculate the exact $\chi$ using the Lehmann representation, 
\begin{equation} \label{chi}
    \chi(x,x',\omega) = \sum_{m} \bigg[ \frac{\langle0|\hat{n}(x)|m\rangle \langle m|\hat{n}(x')|0\rangle}{\omega - (E_{m}-E_{0})+i \eta} + c.c.(-\omega) \bigg],
\end{equation}
where $|0\rangle$, $E_{0}$, $|m\rangle$ and $E_{m}$ are the ground state and its energy, and the $m-$th excited state and its energy, respectively, $\hat{n}$ is the density operator in the Heisenberg picture and $\eta$ is a positive infinitesimal. $\chi$ has poles at the excitation energies of the system, $E_{m}-E_{0}$. It is convenient to calculate Im$(\chi)$, with Re$(\chi)$ following from the Kramers-Kronig relations \cite{Kramers_Kronig, Kramers_Kronig2}. As is customary, we replace $\eta$ with a small positive number, to broaden the absorption peaks for ease of viewing.

We then determine the exact KS potential through our reverse-engineering algorithm \cite{iDEA_RE}. From the exact KS orbitals, we calculate the exact non-interacting density-response function,
\begin{equation} \label{chi_0}
    \chi_{0}(x,x',\omega) = \sum_{i, j} (f_{i}-f_{j})\frac{\phi_{i}^{*}(x)\phi_{j}(x)\phi_{j}^{*}(x')\phi_{i}(x')}{\omega - (\varepsilon_{j}-\varepsilon_{i}) + i \eta},
\end{equation}
where the $\phi_{i}$, $\varepsilon_{i}$ are the exact solutions to the Kohn-Sham equations of ground-state DFT, and $f_{i}$ is the Fermi occupation (0 or 1) of $\phi_{i}$. 

The Dyson equation may be manipulated into an expression for $f_{\mathrm{xc}}$,
\begin{equation} \label{fxc}
f_{\mathrm{xc}} = \chi_{0}^{-1} - \chi^{-1} - u, 
\end{equation}
but the inverses of $\chi$ and $\chi_{0}$ are not well defined. For instance, a spatially uniform perturbation of any angular frequency induces no change in density, so both $\chi$ and $\chi_{0}$ have a zero eigenvalue and therefore a zero determinant. To overcome this, we find a pseudo-inverse of $\chi$ using truncated singular-value decomposition, discarding those eigenvectors with eigenvalues smallest in magnitude, which we term the eigenvalue cutoff. This procedure is repeated for $\chi_{0}$, discarding the same number of eigenvectors. From the modified response functions a kernel $f_{\mathrm{xc}}$ is now well defined. We confirm the validity of this procedure by verifying that the calculated $f_{\mathrm{xc}}$, together with the unmodified $\chi_{0}$, closely reproduces the unmodified $\chi$ via the Dyson equation. Additionally, we ensure that the zero-force sum rule is obeyed -- a well known property of the exact $f_{\mathrm{xc}}$ \footnote{See Supplemental Material for more details.}.

\begin{figure}[htbp]      
\centering
\includegraphics[width=1.0\linewidth]{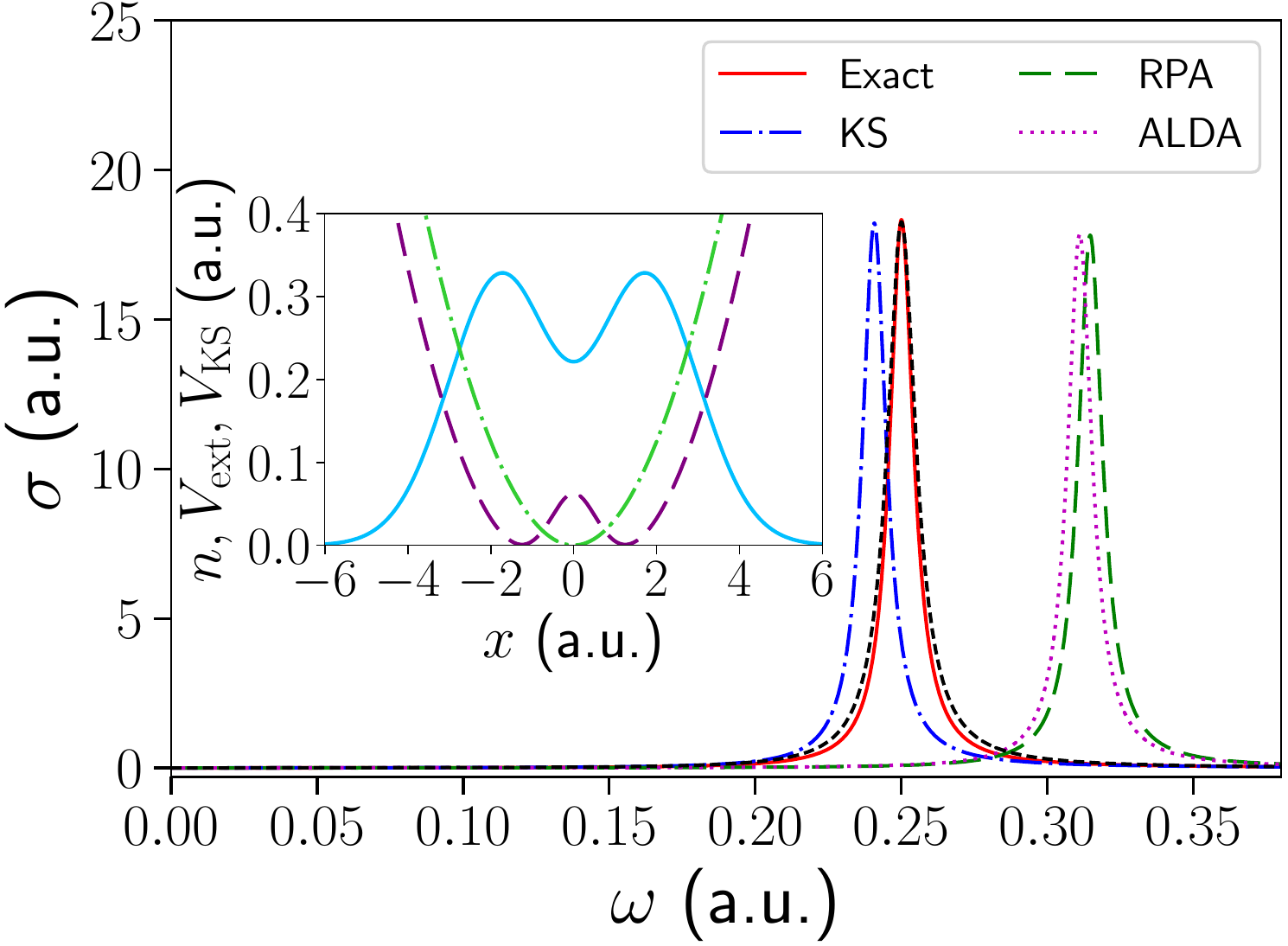}
\caption{Two interacting electrons in a harmonic potential. The inset shows the electron density (solid blue), along with the external (dotted-dashed green) and exact Kohn-Sham (dashed purple) potentials. In the main panel, the absorption spectra (detailing the first excitation) of the exact and Kohn-Sham systems, along with the RPA and ALDA approximations. We check that the calculated $f_{\mathrm{xc}}$ is correct by solving the Dyson equation and comparing the resultant absorption spectrum (short-dashed black) with the exact.}
\label{qho_system}
\end{figure}

We begin by considering a system of two interacting electrons confined to a harmonic well potential ($\omega_{0}=0.25$ a.u.), where $\omega_{0}$ is the angular frequency of the well (inset of Fig.~\ref{qho_system}). We compute the exact optical absorption spectrum of the system \footnote{For this harmonic well system, at the level of linear response theory, only one excitation appears in the absorption spectrum.}, with the first excitation at $\omega = \omega_{0}$ (Fig.~\ref{qho_system}). Additionally, we compute the absorption spectrum of the exact Kohn-Sham system, in which the absorption frequency is slightly too low ($\approx0.01$ a.u.). We also calculate the RPA and ALDA absorption spectra, in which the RPA and ALDA \cite{Entwistle_LDA2} kernels are combined with the \textit{exact} $\chi_{0}$. This last point provides a strong reminder of the challenge of $f_{\mathrm{xc}}$: starting from the exact Kohn-Sham orbitals, a much better absorption peak is obtained by ignoring the induced changes in the Hartree and xc potentials ($\chi_0$) than by accounting for the first exactly and either neglecting (RPA) or approximating (ALDA) the second. This highlights the importance of obtaining a good approximation to the ground-state xc potential $v_{\mathrm{xc}}$, which leads to $\chi_{0}$.

\begin{figure}[htbp]     
\centering
\includegraphics[width=1.0\linewidth]{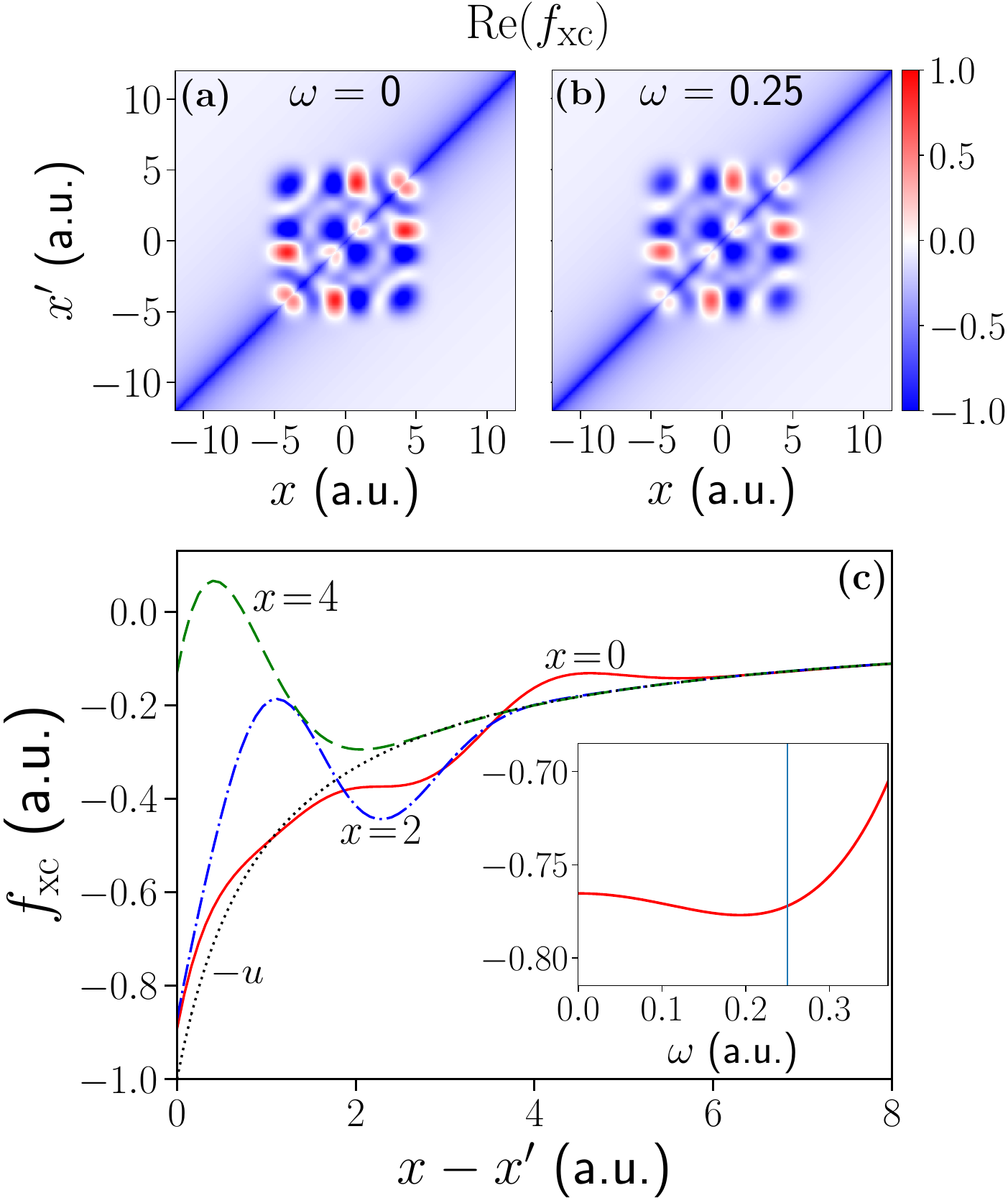}
\caption{The real part of the exact $f_{\mathrm{xc}}$ of the harmonic well system: (a) In the adiabatic limit ($\omega = 0$), and (b) at the first excitation ($\omega = 0.25$). (c) $f_{\mathrm{xc}}$ has a rather simple non-local dependence, which is similar to the negative of the Coulomb interaction $u$. Here we focus on $f_{\mathrm{xc}}$ at $\omega=0.25$. Inset: We observe $f_{\mathrm{xc}}$ to have strikingly weak $\omega$-dependence up to the first excitation (vertical line). We illustrate this by plotting its value (solid red) at a point along the main diagonal ($x=x'$) which corresponds to the peak in electron density in the inset of Fig.~\ref{qho_system} ($x=1.7$ a.u.).}
\label{qho_fxc}
\end{figure}

We now turn to the spatial characteristics of $f_{\mathrm{xc}}$ (Fig.~\ref{qho_fxc}). Typically, several different choices of the eigenvalue cutoff yield kernels $f_{\mathrm{xc}}$ with varying degrees of spatial structure, all of which essentially yield the correct $\chi$ from the exact $\chi_{0}$ as set out above. Of these, we select the eigenvalue cutoff with the largest magnitude, resulting in the smoothest possible spatial structure without detriment to the exact absorption spectrum (Fig.~\ref{qho_system}). We observe that while $f_{\mathrm{xc}}$ has real and imaginary parts (see later), the real part alone is sufficient to reproduce the position and weight of the first excitation $(\omega=\omega_{0})$. Fig.~\ref{qho_fxc}(a) and Fig.~\ref{qho_fxc}(b) show Re$(f_{\mathrm{xc}})$ at $\omega = 0$ and $\omega_{0}$, respectively. The behavior of $f_{\mathrm{xc}}$ away from the diagonal, $x\neq x'$, represents the kernel's non-locality, and it is evident that this is fairly simple in nature; analysis (Fig~\ref{qho_fxc}(c)) shows it to be similar to the negative of the Coulomb interaction, with which it therefore tends to cancel in the expression for $\chi$. The $\omega$-dependence of $f_{\mathrm{xc}}$ up to the first excitation is seen to be extremely weak, as observed in other model systems \cite{fxc_hubbard, fxc_hubbard2, fxc_anharmonic}. We analyze this more closely in the inset to Fig.~\ref{qho_fxc}(c).

\begin{figure}[htbp]     
\centering
\includegraphics[width=1.0\linewidth]{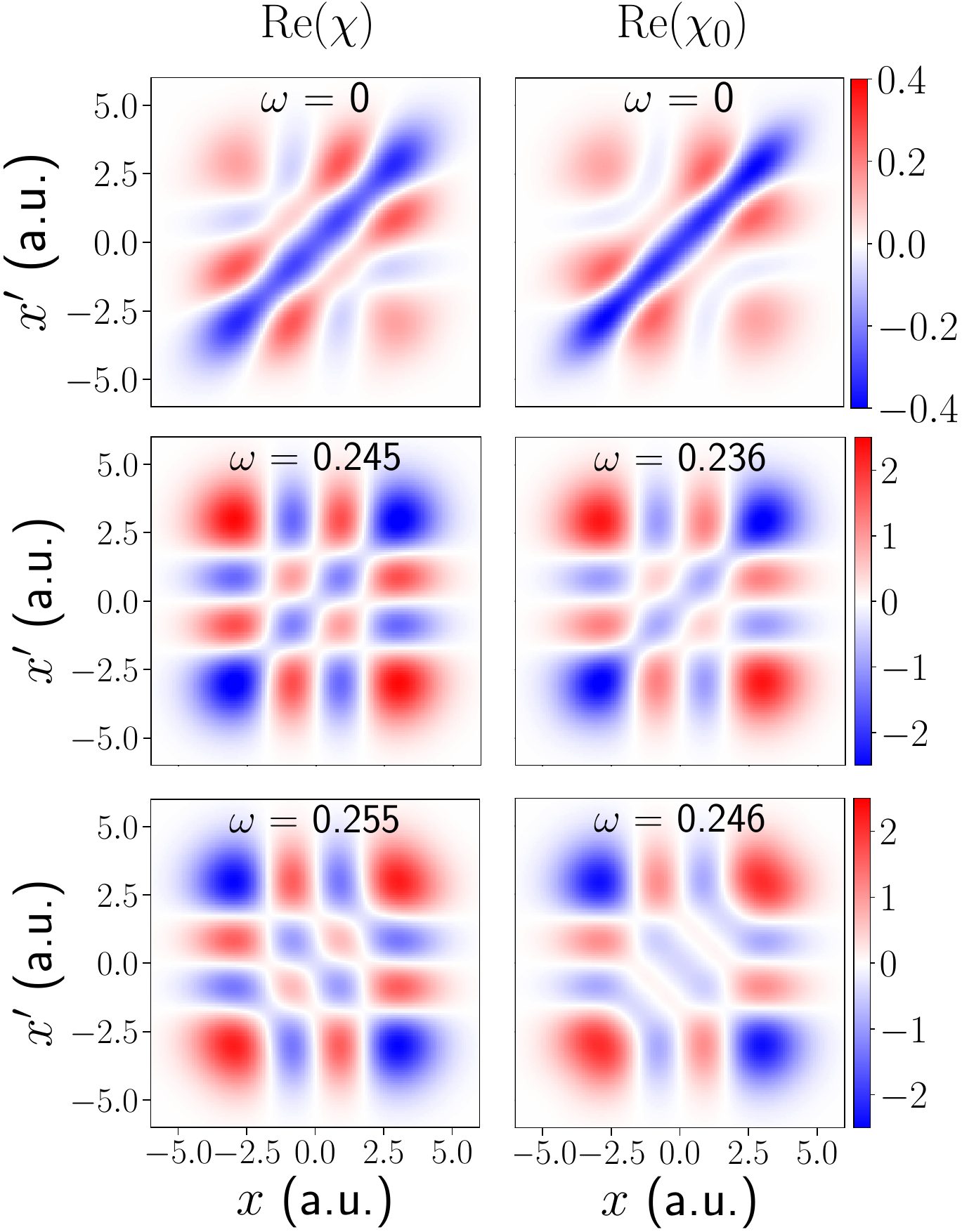}
\caption{The exact $\chi$ and $\chi_{0}$ in the harmonic well system: Left: Re$(\chi)$ at $\omega = 0$ and on either side of the transition at 0.250. Right: Re$(\chi_{0})$ at $\omega = 0$ and on either side of the transition at 0.241. Re$(\chi)$ and Re$(\chi_{0})$ exhibit remarkably similar spatial structure.}
\label{qho_drf}
\end{figure}

To gain insight into these observations, we analyze the exact $\chi$ and $\chi_{0}$. Fig.~\ref{qho_drf} shows Re$(\chi)$ and Re$(\chi_{0})$; up to the first excitation, these exhibit strong -- but closely similar -- $\omega$-dependence. The similarity arises in part from the exact many-electron wavefunction being well approximated by the exact Kohn-Sham wavefunction \footnote{We define this as a Slater determinant of the occupied KS orbitals.}, which reflects the dominance of exchange (including self-interaction correction) in the harmonic potential system \footnote{E. Richardson, private communication}. Therefore $\chi^{-1}$ and $\chi_{0}^{-1}$ largely cancel, so that $f_{\mathrm{xc}}$ is similar to $-u$, with weak $\omega$-dependence.

\begin{figure}[htbp]     
\centering
\includegraphics[width=1.0\linewidth]{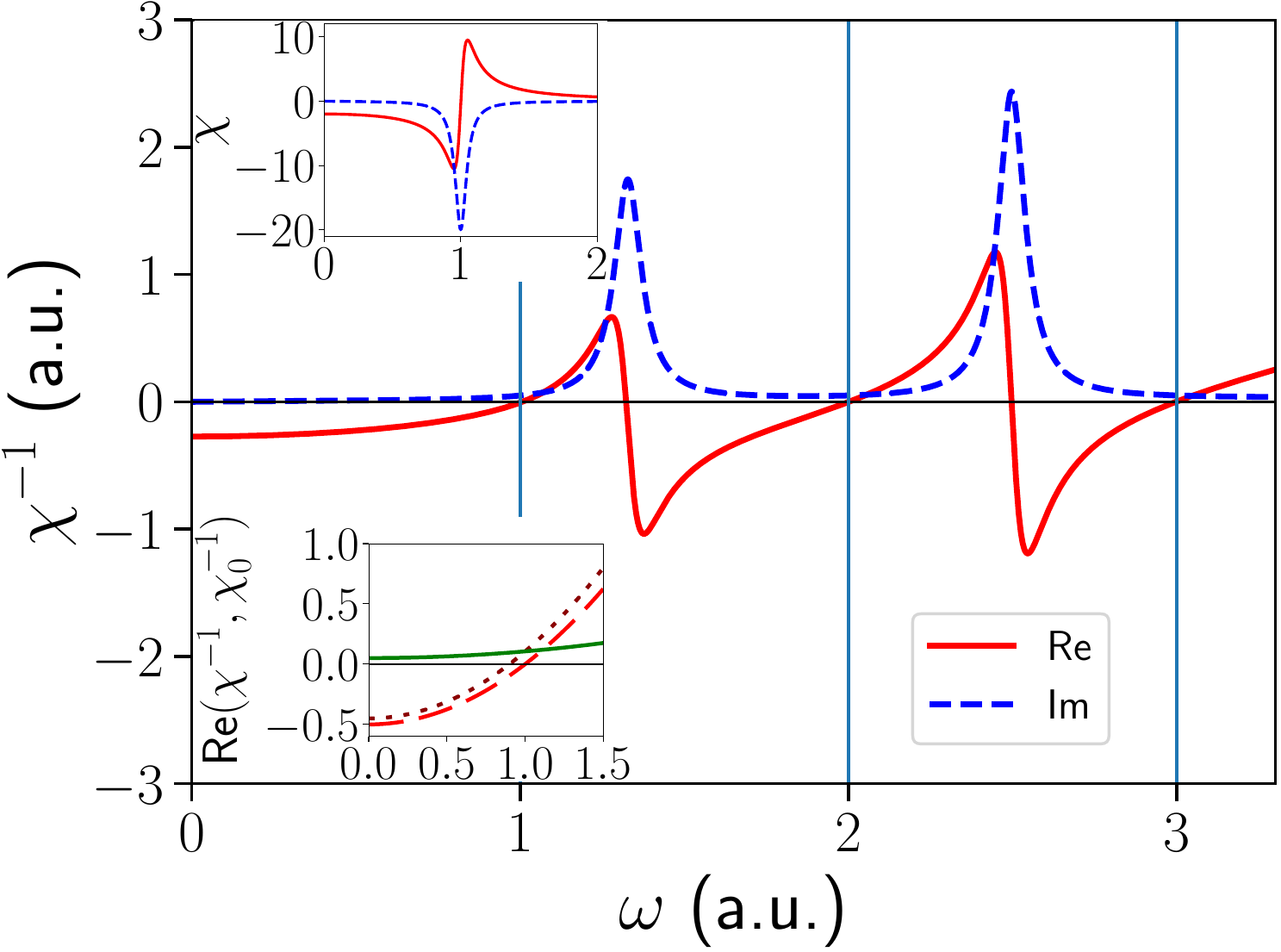}
\caption{The top inset shows $\chi$ for a simple model without spatial dependence and a single excitation at $\omega = 1$; the bottom inset shows the near cancellation (solid green) between Re$(\chi^{-1})$ (dashed red) and Re$(\chi_{0}^{-1})$ (dotted dark red), where $\chi_{0}$ has an excitation at 0.9, causing $f_{\mathrm{xc}}$ to exhibit weak $\omega$-dependence. In the main panel, two further excitations at $\omega = 2$ and 3 have been included, to show that Re$(\chi^{-1})$ passes through zero between excitations, which leads to a non-zero Im$(f_{\mathrm{xc}})$, as does the corresponding feature in $\chi_{0}^{-1}$ (not shown).}
\label{toy_model}
\end{figure}

This can be demonstrated succinctly through a simple model, in which we take Eq.~(\ref{chi}) and Eq.~(\ref{chi_0}), and replace the spatially-dependent numerators (the oscillator strengths) with scalars. Specifically, we consider a system with a single excitation at $\omega = 1$, set the numerator equal to 1, and let $\eta = 0.05$ (top inset of Fig.~\ref{toy_model}). We do the same for the Kohn-Sham system, but choose the excitation to occur at $\omega = 0.9$. By taking their inverses, we calculate Re$(\chi_{0}^{-1}-\chi^{-1})$, which is the $\omega$-dependent part of Re$(f_{\mathrm{xc}})$ in Eq.~(\ref{fxc}), and find this to be small in amplitude and have a fairly weak $\omega$-dependence up to the first excitations (bottom inset of Fig.~\ref{toy_model}). The inclusion of higher excitations, and taking the limit $\eta \rightarrow 0$, change little at these low frequencies. 

Including higher excitations in the model $\chi$ causes Re$(\chi)$ to pass through zero between excitations. At these points Re$(\chi^{-1})$ also passes through zero, and Im$(\chi^{-1})$ peaks sharply (Fig.~\ref{toy_model}). As Im$(f_{\mathrm{xc}})$ = Im$(\chi_{0}^{-1}-\chi^{-1})$, we find that the $f_{\mathrm{xc}}$ in our simple model only has an imaginary component when $\chi$ or $\chi_{0}$ passes through zero between excitations, and hence is completely real up to the first excitations (as $\eta \rightarrow 0$). This supports our finding in the harmonic well system, in which Im$(f_{\mathrm{xc}})$ was very small up to the first excitations, and Re$(f_{\mathrm{xc}})$ was sufficient to reproduce the peak in the absorption spectrum.

\begin{figure}[htbp]      
\centering
\includegraphics[width=1.0\linewidth]{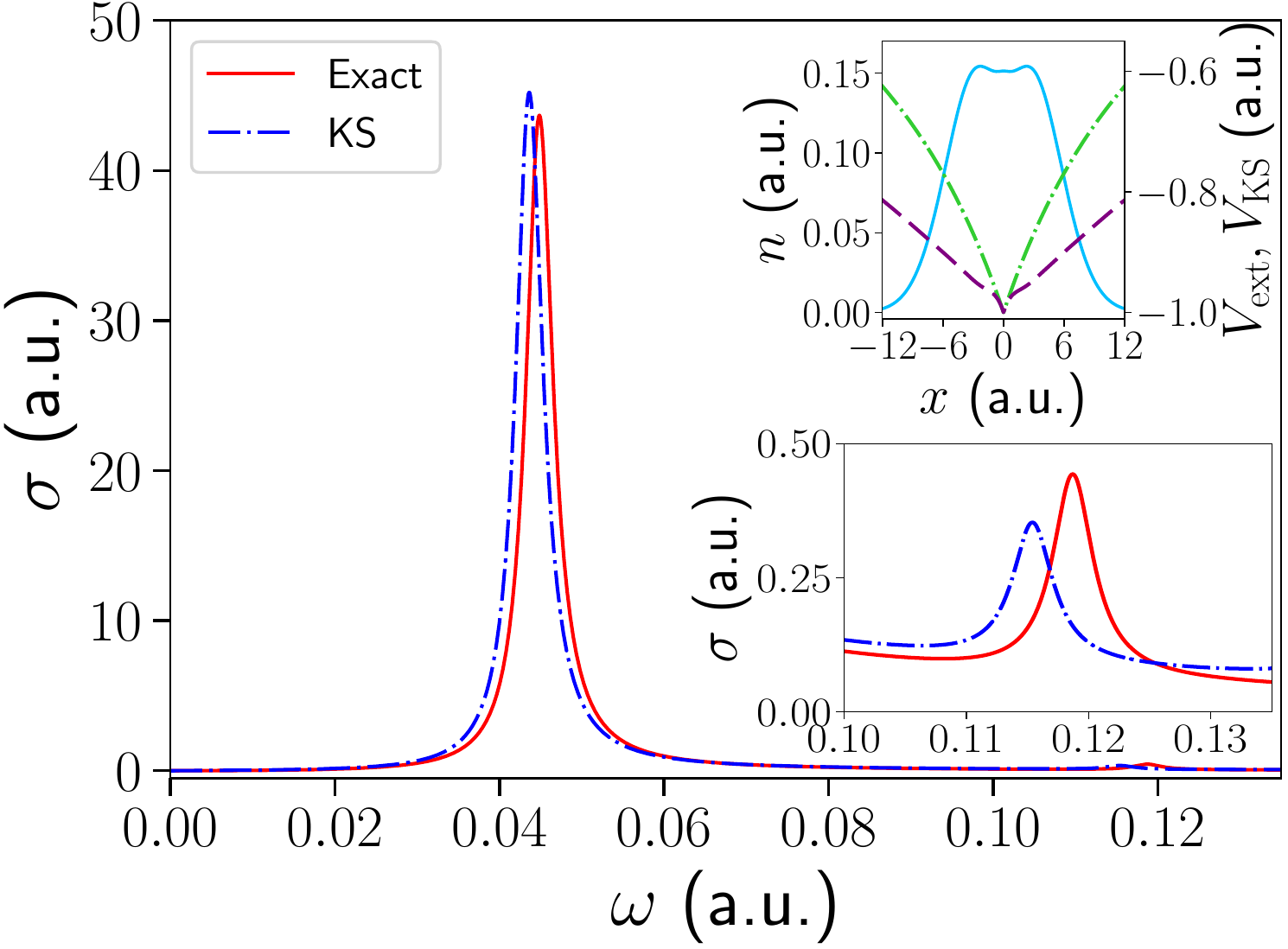}
\caption{Two interacting electrons in an atomic-like potential. The top inset shows the electron density (solid blue), along with the external (dotted-dashed green) and Kohn-Sham (dashed purple) potentials. In the main panel, the absorption spectra of the exact and Kohn-Sham systems; the bottom inset shows the third excitation (fourth in the KS system) in more detail, which is the next to appear after the first excitation.}
\label{atom_system}
\end{figure}

\begin{figure}[htbp]     
\centering
\includegraphics[width=1.0\linewidth]{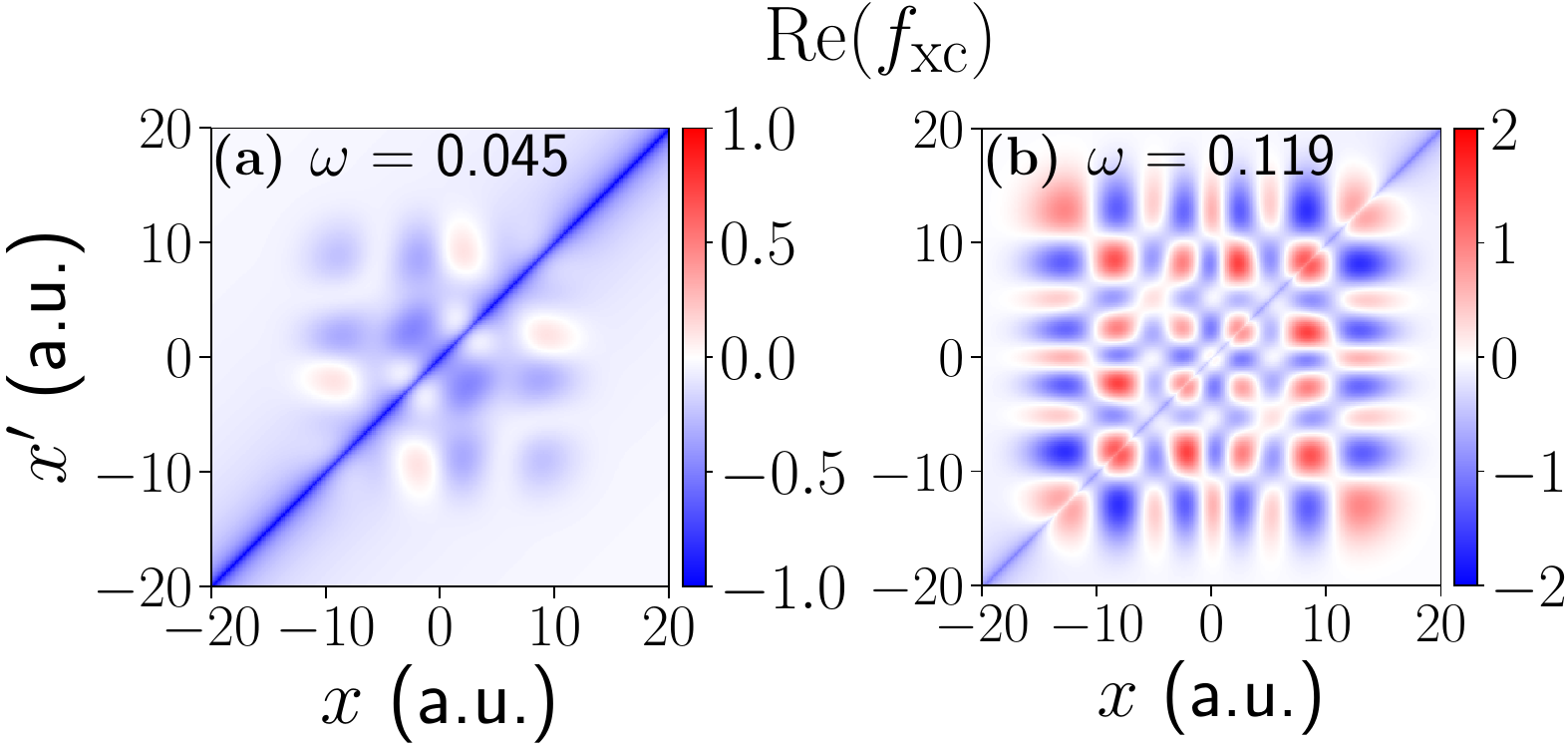}
\caption{The real part of the exact $f_{\mathrm{xc}}$ of the atom-like system: (a) At the first excitation ($\omega = 0.045$), and (b) at the third excitation ($\omega = 0.119$). As in the harmonic well system, we find $f_{\mathrm{xc}}$ to be nearly $\omega$-independent and exhibit a relatively simple spatial form up to the first excitation. However, more complex spatial structure is needed to capture higher excitations.}
\label{atom_fxc}
\end{figure}

We now consider a system whose absorption spectrum includes higher excitations -- two interacting electrons in a softened atomic-like potential (top inset of Fig.~\ref{atom_system}). As in the harmonic well system, the absorption spectrum of the exact Kohn-Sham system is slightly too low for the first excitation (Fig.~\ref{atom_system}). Again, we find $f_{\mathrm{xc}}$ to be dominated by its real part and nearly $\omega$-independent, while exhibiting relatively simple spatial structure, up to and including the first excitation (Fig.~\ref{atom_fxc}(a)). The second excitation does not appear in the absorption spectrum, and so we move to the third excitation, which is much smaller in amplitude than the first, and once more observe the peak in the Kohn-Sham system to be slightly below but still very close to the exact (bottom inset of Fig.~\ref{atom_system}). Again, the closeness between the two peaks arises from the strong similarity between $\chi$ and $\chi_{0}$. In order to reproduce this excitation, a smaller eigenvalue cutoff is required, leading to higher spatial frequencies in $f_{\mathrm{xc}}$ \footnote{As expected for higher energy excited states.} (Fig.~\ref{atom_fxc}(b)).

For the atom-like system, we have investigated the extent to which local kernel approximations for $f_{\mathrm{xc}}$ may be meaningful. As we have observed $f_{\mathrm{xc}}$ to largely cancel with $u$ at low $\omega$, we choose to focus on the Hartree exchange-correlation kernel $f_{\mathrm{Hxc}}=f_{\mathrm{xc}}+u$ which is more local. We incorporate the nonlocal parts of $f_{\mathrm{Hxc}}$ by projecting them onto a local kernel \footnote{We fold the nonlocal $f_{\mathrm{Hxc}}$ with an envelope function that suppresses the more distant nonlocal parts and projects the remainder onto the diagonal $x=x'$.}. We find this largely corrects the difference $\chi_{0}-\chi$, and hence the position of the peak in the absorption spectrum, for the first excitation, but fails to correct the height of the peak. Such a local kernel is completely inadequate to describe the third excitation.

In summary, we have calculated the exact $f_{\mathrm{xc}}(x,x',\omega)$ for two prototype systems. At low $\omega$, we find the imaginary component of $f_{\mathrm{xc}}$ to be small, with the real part alone sufficient to reproduce the first excitation. Up to and including the first excitation, Re$(f_{\mathrm{xc}})$ exhibits strikingly weak $\omega$-dependence, stemming from strong, but closely similar $\omega$-dependence between the interacting and non-interacting density-response functions -- boding well for the applicability of adiabatic kernels. Additionally, Re$(f_{\mathrm{xc}})$ here has a rather simple spatial form, which is similar to the negative of the Coulomb interaction $u$, indicating that approximations to $f_{\mathrm{Hxc}}$ may be more appropriate than those for $f_{\mathrm{xc}}$ alone. For higher excitations, $f_{\mathrm{xc}}$ exhibits both additional spatial structure and stronger $\omega$-dependence, indicating that more sophisticated approximations are needed. Throughout, the absorption spectrum of the exact Kohn-Sham system provides a very good starting point, signifying the crucial importance of an accurate approximation for the ground-state $v_{\mathrm{xc}}$.

We thank Phil Hasnip for helpful discussions. Data created during this research is available from the York Research Database \footnote{M. T. Entwistle and R. W. Godby, http://dx.doi.org/10.15124/56f576ee-b2de-4ca5-9251-831bfc3cae6f (2019).}.
\bibliography{Entwistle_2018}

\begin{thebibliography}{51}%
\makeatletter
\providecommand \@ifxundefined [1]{%
 \@ifx{#1\undefined}
}%
\providecommand \@ifnum [1]{%
 \ifnum #1\expandafter \@firstoftwo
 \else \expandafter \@secondoftwo
 \fi
}%
\providecommand \@ifx [1]{%
 \ifx #1\expandafter \@firstoftwo
 \else \expandafter \@secondoftwo
 \fi
}%
\providecommand \natexlab [1]{#1}%
\providecommand \enquote  [1]{``#1''}%
\providecommand \bibnamefont  [1]{#1}%
\providecommand \bibfnamefont [1]{#1}%
\providecommand \citenamefont [1]{#1}%
\providecommand \href@noop [0]{\@secondoftwo}%
\providecommand \href [0]{\begingroup \@sanitize@url \@href}%
\providecommand \@href[1]{\@@startlink{#1}\@@href}%
\providecommand \@@href[1]{\endgroup#1\@@endlink}%
\providecommand \@sanitize@url [0]{\catcode `\\12\catcode `\$12\catcode
  `\&12\catcode `\#12\catcode `\^12\catcode `\_12\catcode `\%12\relax}%
\providecommand \@@startlink[1]{}%
\providecommand \@@endlink[0]{}%
\providecommand \url  [0]{\begingroup\@sanitize@url \@url }%
\providecommand \@url [1]{\endgroup\@href {#1}{\urlprefix }}%
\providecommand \urlprefix  [0]{URL }%
\providecommand \Eprint [0]{\href }%
\providecommand \doibase [0]{http://dx.doi.org/}%
\providecommand \selectlanguage [0]{\@gobble}%
\providecommand \bibinfo  [0]{\@secondoftwo}%
\providecommand \bibfield  [0]{\@secondoftwo}%
\providecommand \translation [1]{[#1]}%
\providecommand \BibitemOpen [0]{}%
\providecommand \bibitemStop [0]{}%
\providecommand \bibitemNoStop [0]{.\EOS\space}%
\providecommand \EOS [0]{\spacefactor3000\relax}%
\providecommand \BibitemShut  [1]{\csname bibitem#1\endcsname}%
\let\auto@bib@innerbib\@empty
\bibitem [{\citenamefont {Runge}\ and\ \citenamefont {Gross}(1984)}]{TDDFT}%
  \BibitemOpen
  \bibfield  {author} {\bibinfo {author} {\bibfnamefont {E.}~\bibnamefont
  {Runge}}\ and\ \bibinfo {author} {\bibfnamefont {E.~K.~U.}\ \bibnamefont
  {Gross}},\ }\href {\doibase 10.1103/PhysRevLett.52.997} {\bibfield  {journal}
  {\bibinfo  {journal} {Phys. Rev. Lett.}\ }\textbf {\bibinfo {volume} {52}},\
  \bibinfo {pages} {997} (\bibinfo {year} {1984})}\BibitemShut {NoStop}%
\bibitem [{\citenamefont {Gross}\ \emph {et~al.}(1996)\citenamefont {Gross},
  \citenamefont {Dobson},\ and\ \citenamefont {Petersilka}}]{TDDFT3}%
  \BibitemOpen
  \bibfield  {author} {\bibinfo {author} {\bibfnamefont {E.~K.~U.}\
  \bibnamefont {Gross}}, \bibinfo {author} {\bibfnamefont {J.~F.}\ \bibnamefont
  {Dobson}}, \ and\ \bibinfo {author} {\bibfnamefont {M.}~\bibnamefont
  {Petersilka}},\ }\enquote {\bibinfo {title} {Density functional theory of
  time-dependent phenomena},}\ in\ \href {\doibase 10.1007/BFb0016643} {\emph
  {\bibinfo {booktitle} {Density Functional Theory II: Relativistic and Time
  Dependent Extensions}}},\ \bibinfo {editor} {edited by\ \bibinfo {editor}
  {\bibfnamefont {R.~F.}\ \bibnamefont {Nalewajski}}}\ (\bibinfo  {publisher}
  {Springer Berlin Heidelberg},\ \bibinfo {address} {Berlin, Heidelberg},\
  \bibinfo {year} {1996})\ pp.\ \bibinfo {pages} {81--172}\BibitemShut
  {NoStop}%
\bibitem [{Note1()}]{Note1}%
  \BibitemOpen
  \bibinfo {note} {Matrix multiplication for the spatially non-local quantities
  $\chi $, $\chi _{0}$ and $f_{\protect \mathrm {xc}}$, and $\omega
  $-dependence, are implied.}\BibitemShut {Stop}%
\bibitem [{\citenamefont {Petersilka}\ \emph {et~al.}(1996)\citenamefont
  {Petersilka}, \citenamefont {Gossmann},\ and\ \citenamefont {Gross}}]{Dyson}%
  \BibitemOpen
  \bibfield  {author} {\bibinfo {author} {\bibfnamefont {M.}~\bibnamefont
  {Petersilka}}, \bibinfo {author} {\bibfnamefont {U.~J.}\ \bibnamefont
  {Gossmann}}, \ and\ \bibinfo {author} {\bibfnamefont {E.~K.~U.}\ \bibnamefont
  {Gross}},\ }\href {\doibase 10.1103/PhysRevLett.76.1212} {\bibfield
  {journal} {\bibinfo  {journal} {Phys. Rev. Lett.}\ }\textbf {\bibinfo
  {volume} {76}},\ \bibinfo {pages} {1212} (\bibinfo {year}
  {1996})}\BibitemShut {NoStop}%
\bibitem [{\citenamefont {Zangwill}\ and\ \citenamefont
  {Soven}(1980)}]{ALDA_LR}%
  \BibitemOpen
  \bibfield  {author} {\bibinfo {author} {\bibfnamefont {A.}~\bibnamefont
  {Zangwill}}\ and\ \bibinfo {author} {\bibfnamefont {P.}~\bibnamefont
  {Soven}},\ }\href {\doibase 10.1103/PhysRevA.21.1561} {\bibfield  {journal}
  {\bibinfo  {journal} {Phys. Rev. A}\ }\textbf {\bibinfo {volume} {21}},\
  \bibinfo {pages} {1561} (\bibinfo {year} {1980})}\BibitemShut {NoStop}%
\bibitem [{\citenamefont {Gross}\ and\ \citenamefont {Kohn}(1985)}]{ALDA_LR2}%
  \BibitemOpen
  \bibfield  {author} {\bibinfo {author} {\bibfnamefont {E.~K.~U.}\
  \bibnamefont {Gross}}\ and\ \bibinfo {author} {\bibfnamefont
  {W.}~\bibnamefont {Kohn}},\ }\href {\doibase 10.1103/PhysRevLett.55.2850}
  {\bibfield  {journal} {\bibinfo  {journal} {Phys. Rev. Lett.}\ }\textbf
  {\bibinfo {volume} {55}},\ \bibinfo {pages} {2850} (\bibinfo {year}
  {1985})}\BibitemShut {NoStop}%
\bibitem [{\citenamefont {Jamorski}\ \emph {et~al.}(1996)\citenamefont
  {Jamorski}, \citenamefont {Casida},\ and\ \citenamefont
  {Salahub}}]{fxc_alda}%
  \BibitemOpen
  \bibfield  {author} {\bibinfo {author} {\bibfnamefont {C.}~\bibnamefont
  {Jamorski}}, \bibinfo {author} {\bibfnamefont {M.~E.}\ \bibnamefont
  {Casida}}, \ and\ \bibinfo {author} {\bibfnamefont {D.~R.}\ \bibnamefont
  {Salahub}},\ }\href {\doibase 10.1063/1.471140} {\bibfield  {journal}
  {\bibinfo  {journal} {The Journal of Chemical Physics}\ }\textbf {\bibinfo
  {volume} {104}},\ \bibinfo {pages} {5134} (\bibinfo {year} {1996})},\ \Eprint
  {http://arxiv.org/abs/https://doi.org/10.1063/1.471140}
  {https://doi.org/10.1063/1.471140} \BibitemShut {NoStop}%
\bibitem [{\citenamefont {Maitra}\ \emph {et~al.}(2004)\citenamefont {Maitra},
  \citenamefont {Zhang}, \citenamefont {Cave},\ and\ \citenamefont
  {Burke}}]{fxc_alda2}%
  \BibitemOpen
  \bibfield  {author} {\bibinfo {author} {\bibfnamefont {N.~T.}\ \bibnamefont
  {Maitra}}, \bibinfo {author} {\bibfnamefont {F.}~\bibnamefont {Zhang}},
  \bibinfo {author} {\bibfnamefont {R.~J.}\ \bibnamefont {Cave}}, \ and\
  \bibinfo {author} {\bibfnamefont {K.}~\bibnamefont {Burke}},\ }\href
  {\doibase 10.1063/1.1651060} {\bibfield  {journal} {\bibinfo  {journal} {The
  Journal of Chemical Physics}\ }\textbf {\bibinfo {volume} {120}},\ \bibinfo
  {pages} {5932} (\bibinfo {year} {2004})},\ \Eprint
  {http://arxiv.org/abs/https://doi.org/10.1063/1.1651060}
  {https://doi.org/10.1063/1.1651060} \BibitemShut {NoStop}%
\bibitem [{\citenamefont {Gavrilenko}\ and\ \citenamefont
  {Bechstedt}(1997)}]{fxc_rpa_lda}%
  \BibitemOpen
  \bibfield  {author} {\bibinfo {author} {\bibfnamefont {V.~I.}\ \bibnamefont
  {Gavrilenko}}\ and\ \bibinfo {author} {\bibfnamefont {F.}~\bibnamefont
  {Bechstedt}},\ }\href {\doibase 10.1103/PhysRevB.55.4343} {\bibfield
  {journal} {\bibinfo  {journal} {Phys. Rev. B}\ }\textbf {\bibinfo {volume}
  {55}},\ \bibinfo {pages} {4343} (\bibinfo {year} {1997})}\BibitemShut
  {NoStop}%
\bibitem [{\citenamefont {Kim}\ and\ \citenamefont
  {G\"orling}(2002{\natexlab{a}})}]{fxc_exx}%
  \BibitemOpen
  \bibfield  {author} {\bibinfo {author} {\bibfnamefont {Y.-H.}\ \bibnamefont
  {Kim}}\ and\ \bibinfo {author} {\bibfnamefont {A.}~\bibnamefont
  {G\"orling}},\ }\href {\doibase 10.1103/PhysRevLett.89.096402} {\bibfield
  {journal} {\bibinfo  {journal} {Phys. Rev. Lett.}\ }\textbf {\bibinfo
  {volume} {89}},\ \bibinfo {pages} {096402} (\bibinfo {year}
  {2002}{\natexlab{a}})}\BibitemShut {NoStop}%
\bibitem [{\citenamefont {Kim}\ and\ \citenamefont
  {G\"orling}(2002{\natexlab{b}})}]{fxc_exx2}%
  \BibitemOpen
  \bibfield  {author} {\bibinfo {author} {\bibfnamefont {Y.-H.}\ \bibnamefont
  {Kim}}\ and\ \bibinfo {author} {\bibfnamefont {A.}~\bibnamefont
  {G\"orling}},\ }\href {\doibase 10.1103/PhysRevB.66.035114} {\bibfield
  {journal} {\bibinfo  {journal} {Phys. Rev. B}\ }\textbf {\bibinfo {volume}
  {66}},\ \bibinfo {pages} {035114} (\bibinfo {year}
  {2002}{\natexlab{b}})}\BibitemShut {NoStop}%
\bibitem [{\citenamefont {G\"orling}(1998)}]{fxc_exx3}%
  \BibitemOpen
  \bibfield  {author} {\bibinfo {author} {\bibfnamefont {A.}~\bibnamefont
  {G\"orling}},\ }\href {\doibase 10.1103/PhysRevA.57.3433} {\bibfield
  {journal} {\bibinfo  {journal} {Phys. Rev. A}\ }\textbf {\bibinfo {volume}
  {57}},\ \bibinfo {pages} {3433} (\bibinfo {year} {1998})}\BibitemShut
  {NoStop}%
\bibitem [{\citenamefont {Hellgren}\ and\ \citenamefont {von
  Barth}(2009)}]{fxc_exx4}%
  \BibitemOpen
  \bibfield  {author} {\bibinfo {author} {\bibfnamefont {M.}~\bibnamefont
  {Hellgren}}\ and\ \bibinfo {author} {\bibfnamefont {U.}~\bibnamefont {von
  Barth}},\ }\href {\doibase 10.1063/1.3179756} {\bibfield  {journal} {\bibinfo
   {journal} {The Journal of Chemical Physics}\ }\textbf {\bibinfo {volume}
  {131}},\ \bibinfo {pages} {044110} (\bibinfo {year} {2009})},\ \Eprint
  {http://arxiv.org/abs/https://doi.org/10.1063/1.3179756}
  {https://doi.org/10.1063/1.3179756} \BibitemShut {NoStop}%
\bibitem [{\citenamefont {Ipatov}\ \emph {et~al.}(2010)\citenamefont {Ipatov},
  \citenamefont {Heßelmann},\ and\ \citenamefont {Görling}}]{fxc_exx5}%
  \BibitemOpen
  \bibfield  {author} {\bibinfo {author} {\bibfnamefont {A.}~\bibnamefont
  {Ipatov}}, \bibinfo {author} {\bibfnamefont {A.}~\bibnamefont {Heßelmann}},
  \ and\ \bibinfo {author} {\bibfnamefont {A.}~\bibnamefont {Görling}},\
  }\href {\doibase 10.1002/qua.22561} {\bibfield  {journal} {\bibinfo
  {journal} {International Journal of Quantum Chemistry}\ }\textbf {\bibinfo
  {volume} {110}},\ \bibinfo {pages} {2202} (\bibinfo {year} {2010})},\ \Eprint
  {http://arxiv.org/abs/https://onlinelibrary.wiley.com/doi/pdf/10.1002/qua.22561}
  {https://onlinelibrary.wiley.com/doi/pdf/10.1002/qua.22561} \BibitemShut
  {NoStop}%
\bibitem [{\citenamefont {G\"orling}(1999)}]{fxc_exx6}%
  \BibitemOpen
  \bibfield  {author} {\bibinfo {author} {\bibfnamefont {A.}~\bibnamefont
  {G\"orling}},\ }\href {\doibase 10.1103/PhysRevLett.83.5459} {\bibfield
  {journal} {\bibinfo  {journal} {Phys. Rev. Lett.}\ }\textbf {\bibinfo
  {volume} {83}},\ \bibinfo {pages} {5459} (\bibinfo {year}
  {1999})}\BibitemShut {NoStop}%
\bibitem [{\citenamefont {Tokatly}\ and\ \citenamefont
  {Pankratov}(2001)}]{fxc_pert}%
  \BibitemOpen
  \bibfield  {author} {\bibinfo {author} {\bibfnamefont {I.~V.}\ \bibnamefont
  {Tokatly}}\ and\ \bibinfo {author} {\bibfnamefont {O.}~\bibnamefont
  {Pankratov}},\ }\href {\doibase 10.1103/PhysRevLett.86.2078} {\bibfield
  {journal} {\bibinfo  {journal} {Phys. Rev. Lett.}\ }\textbf {\bibinfo
  {volume} {86}},\ \bibinfo {pages} {2078} (\bibinfo {year}
  {2001})}\BibitemShut {NoStop}%
\bibitem [{\citenamefont {Tokatly}\ \emph {et~al.}(2002)\citenamefont
  {Tokatly}, \citenamefont {Stubner},\ and\ \citenamefont
  {Pankratov}}]{fxc_pert2}%
  \BibitemOpen
  \bibfield  {author} {\bibinfo {author} {\bibfnamefont {I.~V.}\ \bibnamefont
  {Tokatly}}, \bibinfo {author} {\bibfnamefont {R.}~\bibnamefont {Stubner}}, \
  and\ \bibinfo {author} {\bibfnamefont {O.}~\bibnamefont {Pankratov}},\ }\href
  {\doibase 10.1103/PhysRevB.65.113107} {\bibfield  {journal} {\bibinfo
  {journal} {Phys. Rev. B}\ }\textbf {\bibinfo {volume} {65}},\ \bibinfo
  {pages} {113107} (\bibinfo {year} {2002})}\BibitemShut {NoStop}%
\bibitem [{\citenamefont {Botti}\ \emph {et~al.}(2004)\citenamefont {Botti},
  \citenamefont {Sottile}, \citenamefont {Vast}, \citenamefont {Olevano},
  \citenamefont {Reining}, \citenamefont {Weissker}, \citenamefont {Rubio},
  \citenamefont {Onida}, \citenamefont {Del~Sole},\ and\ \citenamefont
  {Godby}}]{fxc_nonlocal}%
  \BibitemOpen
  \bibfield  {author} {\bibinfo {author} {\bibfnamefont {S.}~\bibnamefont
  {Botti}}, \bibinfo {author} {\bibfnamefont {F.}~\bibnamefont {Sottile}},
  \bibinfo {author} {\bibfnamefont {N.}~\bibnamefont {Vast}}, \bibinfo {author}
  {\bibfnamefont {V.}~\bibnamefont {Olevano}}, \bibinfo {author} {\bibfnamefont
  {L.}~\bibnamefont {Reining}}, \bibinfo {author} {\bibfnamefont {H.-C.}\
  \bibnamefont {Weissker}}, \bibinfo {author} {\bibfnamefont {A.}~\bibnamefont
  {Rubio}}, \bibinfo {author} {\bibfnamefont {G.}~\bibnamefont {Onida}},
  \bibinfo {author} {\bibfnamefont {R.}~\bibnamefont {Del~Sole}}, \ and\
  \bibinfo {author} {\bibfnamefont {R.~W.}\ \bibnamefont {Godby}},\ }\href
  {\doibase 10.1103/PhysRevB.69.155112} {\bibfield  {journal} {\bibinfo
  {journal} {Phys. Rev. B}\ }\textbf {\bibinfo {volume} {69}},\ \bibinfo
  {pages} {155112} (\bibinfo {year} {2004})}\BibitemShut {NoStop}%
\bibitem [{\citenamefont {Sharma}\ \emph {et~al.}(2011)\citenamefont {Sharma},
  \citenamefont {Dewhurst}, \citenamefont {Sanna},\ and\ \citenamefont
  {Gross}}]{fxc_nonlocal2}%
  \BibitemOpen
  \bibfield  {author} {\bibinfo {author} {\bibfnamefont {S.}~\bibnamefont
  {Sharma}}, \bibinfo {author} {\bibfnamefont {J.~K.}\ \bibnamefont
  {Dewhurst}}, \bibinfo {author} {\bibfnamefont {A.}~\bibnamefont {Sanna}}, \
  and\ \bibinfo {author} {\bibfnamefont {E.~K.~U.}\ \bibnamefont {Gross}},\
  }\href {\doibase 10.1103/PhysRevLett.107.186401} {\bibfield  {journal}
  {\bibinfo  {journal} {Phys. Rev. Lett.}\ }\textbf {\bibinfo {volume} {107}},\
  \bibinfo {pages} {186401} (\bibinfo {year} {2011})}\BibitemShut {NoStop}%
\bibitem [{\citenamefont {Trevisanutto}\ \emph {et~al.}(2013)\citenamefont
  {Trevisanutto}, \citenamefont {Terentjevs}, \citenamefont {Constantin},
  \citenamefont {Olevano},\ and\ \citenamefont {Sala}}]{fxc_nonlocal3}%
  \BibitemOpen
  \bibfield  {author} {\bibinfo {author} {\bibfnamefont {P.~E.}\ \bibnamefont
  {Trevisanutto}}, \bibinfo {author} {\bibfnamefont {A.}~\bibnamefont
  {Terentjevs}}, \bibinfo {author} {\bibfnamefont {L.~A.}\ \bibnamefont
  {Constantin}}, \bibinfo {author} {\bibfnamefont {V.}~\bibnamefont {Olevano}},
  \ and\ \bibinfo {author} {\bibfnamefont {F.~D.}\ \bibnamefont {Sala}},\
  }\href {\doibase 10.1103/PhysRevB.87.205143} {\bibfield  {journal} {\bibinfo
  {journal} {Phys. Rev. B}\ }\textbf {\bibinfo {volume} {87}},\ \bibinfo
  {pages} {205143} (\bibinfo {year} {2013})}\BibitemShut {NoStop}%
\bibitem [{\citenamefont {Rigamonti}\ \emph {et~al.}(2015)\citenamefont
  {Rigamonti}, \citenamefont {Botti}, \citenamefont {Veniard}, \citenamefont
  {Draxl}, \citenamefont {Reining},\ and\ \citenamefont
  {Sottile}}]{fxc_nonlocal4}%
  \BibitemOpen
  \bibfield  {author} {\bibinfo {author} {\bibfnamefont {S.}~\bibnamefont
  {Rigamonti}}, \bibinfo {author} {\bibfnamefont {S.}~\bibnamefont {Botti}},
  \bibinfo {author} {\bibfnamefont {V.}~\bibnamefont {Veniard}}, \bibinfo
  {author} {\bibfnamefont {C.}~\bibnamefont {Draxl}}, \bibinfo {author}
  {\bibfnamefont {L.}~\bibnamefont {Reining}}, \ and\ \bibinfo {author}
  {\bibfnamefont {F.}~\bibnamefont {Sottile}},\ }\href {\doibase
  10.1103/PhysRevLett.114.146402} {\bibfield  {journal} {\bibinfo  {journal}
  {Phys. Rev. Lett.}\ }\textbf {\bibinfo {volume} {114}},\ \bibinfo {pages}
  {146402} (\bibinfo {year} {2015})}\BibitemShut {NoStop}%
\bibitem [{\citenamefont {Nazarov}\ \emph {et~al.}(2010)\citenamefont
  {Nazarov}, \citenamefont {Tokatly}, \citenamefont {Pittalis},\ and\
  \citenamefont {Vignale}}]{fxc_heg}%
  \BibitemOpen
  \bibfield  {author} {\bibinfo {author} {\bibfnamefont {V.~U.}\ \bibnamefont
  {Nazarov}}, \bibinfo {author} {\bibfnamefont {I.~V.}\ \bibnamefont
  {Tokatly}}, \bibinfo {author} {\bibfnamefont {S.}~\bibnamefont {Pittalis}}, \
  and\ \bibinfo {author} {\bibfnamefont {G.}~\bibnamefont {Vignale}},\ }\href
  {\doibase 10.1103/PhysRevB.81.245101} {\bibfield  {journal} {\bibinfo
  {journal} {Phys. Rev. B}\ }\textbf {\bibinfo {volume} {81}},\ \bibinfo
  {pages} {245101} (\bibinfo {year} {2010})}\BibitemShut {NoStop}%
\bibitem [{\citenamefont {Qian}\ and\ \citenamefont
  {Vignale}(2002)}]{fxc_heg2}%
  \BibitemOpen
  \bibfield  {author} {\bibinfo {author} {\bibfnamefont {Z.}~\bibnamefont
  {Qian}}\ and\ \bibinfo {author} {\bibfnamefont {G.}~\bibnamefont {Vignale}},\
  }\href {\doibase 10.1103/PhysRevB.65.235121} {\bibfield  {journal} {\bibinfo
  {journal} {Phys. Rev. B}\ }\textbf {\bibinfo {volume} {65}},\ \bibinfo
  {pages} {235121} (\bibinfo {year} {2002})}\BibitemShut {NoStop}%
\bibitem [{\citenamefont {Richardson}\ and\ \citenamefont
  {Ashcroft}(1994)}]{fxc_heg3}%
  \BibitemOpen
  \bibfield  {author} {\bibinfo {author} {\bibfnamefont {C.~F.}\ \bibnamefont
  {Richardson}}\ and\ \bibinfo {author} {\bibfnamefont {N.~W.}\ \bibnamefont
  {Ashcroft}},\ }\href {\doibase 10.1103/PhysRevB.50.8170} {\bibfield
  {journal} {\bibinfo  {journal} {Phys. Rev. B}\ }\textbf {\bibinfo {volume}
  {50}},\ \bibinfo {pages} {8170} (\bibinfo {year} {1994})}\BibitemShut
  {NoStop}%
\bibitem [{\citenamefont {Conti}\ \emph {et~al.}(1997)\citenamefont {Conti},
  \citenamefont {Nifosì},\ and\ \citenamefont {Tosi}}]{fxc_heg4}%
  \BibitemOpen
  \bibfield  {author} {\bibinfo {author} {\bibfnamefont {S.}~\bibnamefont
  {Conti}}, \bibinfo {author} {\bibfnamefont {R.}~\bibnamefont {Nifosì}}, \
  and\ \bibinfo {author} {\bibfnamefont {M.~P.}\ \bibnamefont {Tosi}},\ }\href
  {http://stacks.iop.org/0953-8984/9/i=34/a=004} {\bibfield  {journal}
  {\bibinfo  {journal} {Journal of Physics: Condensed Matter}\ }\textbf
  {\bibinfo {volume} {9}},\ \bibinfo {pages} {L475} (\bibinfo {year}
  {1997})}\BibitemShut {NoStop}%
\bibitem [{\citenamefont {Panholzer}\ \emph {et~al.}(2018)\citenamefont
  {Panholzer}, \citenamefont {Gatti},\ and\ \citenamefont
  {Reining}}]{fxc_heg5}%
  \BibitemOpen
  \bibfield  {author} {\bibinfo {author} {\bibfnamefont {M.}~\bibnamefont
  {Panholzer}}, \bibinfo {author} {\bibfnamefont {M.}~\bibnamefont {Gatti}}, \
  and\ \bibinfo {author} {\bibfnamefont {L.}~\bibnamefont {Reining}},\ }\href
  {\doibase 10.1103/PhysRevLett.120.166402} {\bibfield  {journal} {\bibinfo
  {journal} {Phys. Rev. Lett.}\ }\textbf {\bibinfo {volume} {120}},\ \bibinfo
  {pages} {166402} (\bibinfo {year} {2018})}\BibitemShut {NoStop}%
\bibitem [{\citenamefont {Onida}\ \emph {et~al.}(2002)\citenamefont {Onida},
  \citenamefont {Reining},\ and\ \citenamefont {Rubio}}]{fxc_bse}%
  \BibitemOpen
  \bibfield  {author} {\bibinfo {author} {\bibfnamefont {G.}~\bibnamefont
  {Onida}}, \bibinfo {author} {\bibfnamefont {L.}~\bibnamefont {Reining}}, \
  and\ \bibinfo {author} {\bibfnamefont {A.}~\bibnamefont {Rubio}},\ }\href
  {\doibase 10.1103/RevModPhys.74.601} {\bibfield  {journal} {\bibinfo
  {journal} {Rev. Mod. Phys.}\ }\textbf {\bibinfo {volume} {74}},\ \bibinfo
  {pages} {601} (\bibinfo {year} {2002})}\BibitemShut {NoStop}%
\bibitem [{\citenamefont {Reining}\ \emph {et~al.}(2002)\citenamefont
  {Reining}, \citenamefont {Olevano}, \citenamefont {Rubio},\ and\
  \citenamefont {Onida}}]{fxc_bse2}%
  \BibitemOpen
  \bibfield  {author} {\bibinfo {author} {\bibfnamefont {L.}~\bibnamefont
  {Reining}}, \bibinfo {author} {\bibfnamefont {V.}~\bibnamefont {Olevano}},
  \bibinfo {author} {\bibfnamefont {A.}~\bibnamefont {Rubio}}, \ and\ \bibinfo
  {author} {\bibfnamefont {G.}~\bibnamefont {Onida}},\ }\href {\doibase
  10.1103/PhysRevLett.88.066404} {\bibfield  {journal} {\bibinfo  {journal}
  {Phys. Rev. Lett.}\ }\textbf {\bibinfo {volume} {88}},\ \bibinfo {pages}
  {066404} (\bibinfo {year} {2002})}\BibitemShut {NoStop}%
\bibitem [{\citenamefont {Sottile}\ \emph {et~al.}(2003)\citenamefont
  {Sottile}, \citenamefont {Olevano},\ and\ \citenamefont
  {Reining}}]{fxc_bse3}%
  \BibitemOpen
  \bibfield  {author} {\bibinfo {author} {\bibfnamefont {F.}~\bibnamefont
  {Sottile}}, \bibinfo {author} {\bibfnamefont {V.}~\bibnamefont {Olevano}}, \
  and\ \bibinfo {author} {\bibfnamefont {L.}~\bibnamefont {Reining}},\ }\href
  {\doibase 10.1103/PhysRevLett.91.056402} {\bibfield  {journal} {\bibinfo
  {journal} {Phys. Rev. Lett.}\ }\textbf {\bibinfo {volume} {91}},\ \bibinfo
  {pages} {056402} (\bibinfo {year} {2003})}\BibitemShut {NoStop}%
\bibitem [{\citenamefont {Adragna}\ \emph {et~al.}(2003)\citenamefont
  {Adragna}, \citenamefont {Del~Sole},\ and\ \citenamefont
  {Marini}}]{fxc_bse4}%
  \BibitemOpen
  \bibfield  {author} {\bibinfo {author} {\bibfnamefont {G.}~\bibnamefont
  {Adragna}}, \bibinfo {author} {\bibfnamefont {R.}~\bibnamefont {Del~Sole}}, \
  and\ \bibinfo {author} {\bibfnamefont {A.}~\bibnamefont {Marini}},\ }\href
  {\doibase 10.1103/PhysRevB.68.165108} {\bibfield  {journal} {\bibinfo
  {journal} {Phys. Rev. B}\ }\textbf {\bibinfo {volume} {68}},\ \bibinfo
  {pages} {165108} (\bibinfo {year} {2003})}\BibitemShut {NoStop}%
\bibitem [{\citenamefont {Marini}\ \emph {et~al.}(2003)\citenamefont {Marini},
  \citenamefont {Del~Sole},\ and\ \citenamefont {Rubio}}]{fxc_bse5}%
  \BibitemOpen
  \bibfield  {author} {\bibinfo {author} {\bibfnamefont {A.}~\bibnamefont
  {Marini}}, \bibinfo {author} {\bibfnamefont {R.}~\bibnamefont {Del~Sole}}, \
  and\ \bibinfo {author} {\bibfnamefont {A.}~\bibnamefont {Rubio}},\ }\href
  {\doibase 10.1103/PhysRevLett.91.256402} {\bibfield  {journal} {\bibinfo
  {journal} {Phys. Rev. Lett.}\ }\textbf {\bibinfo {volume} {91}},\ \bibinfo
  {pages} {256402} (\bibinfo {year} {2003})}\BibitemShut {NoStop}%
\bibitem [{\citenamefont {Thiele}\ and\ \citenamefont
  {K\"ummel}(2009)}]{fxc_sumrule2}%
  \BibitemOpen
  \bibfield  {author} {\bibinfo {author} {\bibfnamefont {M.}~\bibnamefont
  {Thiele}}\ and\ \bibinfo {author} {\bibfnamefont {S.}~\bibnamefont
  {K\"ummel}},\ }\href {\doibase 10.1103/PhysRevA.80.012514} {\bibfield
  {journal} {\bibinfo  {journal} {Phys. Rev. A}\ }\textbf {\bibinfo {volume}
  {80}},\ \bibinfo {pages} {012514} (\bibinfo {year} {2009})}\BibitemShut
  {NoStop}%
\bibitem [{\citenamefont {Aryasetiawan}\ and\ \citenamefont
  {Gunnarsson}(2002)}]{fxc_hubbard}%
  \BibitemOpen
  \bibfield  {author} {\bibinfo {author} {\bibfnamefont {F.}~\bibnamefont
  {Aryasetiawan}}\ and\ \bibinfo {author} {\bibfnamefont {O.}~\bibnamefont
  {Gunnarsson}},\ }\href {\doibase 10.1103/PhysRevB.66.165119} {\bibfield
  {journal} {\bibinfo  {journal} {Phys. Rev. B}\ }\textbf {\bibinfo {volume}
  {66}},\ \bibinfo {pages} {165119} (\bibinfo {year} {2002})}\BibitemShut
  {NoStop}%
\bibitem [{\citenamefont {Carrascal}\ \emph {et~al.}(2018)\citenamefont
  {Carrascal}, \citenamefont {Ferrer}, \citenamefont {Maitra},\ and\
  \citenamefont {Burke}}]{fxc_hubbard2}%
  \BibitemOpen
  \bibfield  {author} {\bibinfo {author} {\bibfnamefont {D.~J.}\ \bibnamefont
  {Carrascal}}, \bibinfo {author} {\bibfnamefont {J.}~\bibnamefont {Ferrer}},
  \bibinfo {author} {\bibfnamefont {N.}~\bibnamefont {Maitra}}, \ and\ \bibinfo
  {author} {\bibfnamefont {K.}~\bibnamefont {Burke}},\ }\href {\doibase
  10.1140/epjb/e2018-90114-9} {\bibfield  {journal} {\bibinfo  {journal} {The
  European Physical Journal B}\ }\textbf {\bibinfo {volume} {91}},\ \bibinfo
  {pages} {142} (\bibinfo {year} {2018})}\BibitemShut {NoStop}%
\bibitem [{\citenamefont {Thiele}\ and\ \citenamefont
  {K\"ummel}(2014)}]{fxc_anharmonic}%
  \BibitemOpen
  \bibfield  {author} {\bibinfo {author} {\bibfnamefont {M.}~\bibnamefont
  {Thiele}}\ and\ \bibinfo {author} {\bibfnamefont {S.}~\bibnamefont
  {K\"ummel}},\ }\href {\doibase 10.1103/PhysRevLett.112.083001} {\bibfield
  {journal} {\bibinfo  {journal} {Phys. Rev. Lett.}\ }\textbf {\bibinfo
  {volume} {112}},\ \bibinfo {pages} {083001} (\bibinfo {year}
  {2014})}\BibitemShut {NoStop}%
\bibitem [{\citenamefont {Maitra}\ and\ \citenamefont
  {Tempel}(2006)}]{fxc_longrange}%
  \BibitemOpen
  \bibfield  {author} {\bibinfo {author} {\bibfnamefont {N.~T.}\ \bibnamefont
  {Maitra}}\ and\ \bibinfo {author} {\bibfnamefont {D.~G.}\ \bibnamefont
  {Tempel}},\ }\href {\doibase 10.1063/1.2387951} {\bibfield  {journal}
  {\bibinfo  {journal} {The Journal of Chemical Physics}\ }\textbf {\bibinfo
  {volume} {125}},\ \bibinfo {pages} {184111} (\bibinfo {year} {2006})},\
  \Eprint {http://arxiv.org/abs/https://doi.org/10.1063/1.2387951}
  {https://doi.org/10.1063/1.2387951} \BibitemShut {NoStop}%
\bibitem [{\citenamefont {Hodgson}\ \emph {et~al.}(2013)\citenamefont
  {Hodgson}, \citenamefont {Ramsden}, \citenamefont {Chapman}, \citenamefont
  {Lillystone},\ and\ \citenamefont {Godby}}]{iDEA}%
  \BibitemOpen
  \bibfield  {author} {\bibinfo {author} {\bibfnamefont {M.~J.~P.}\
  \bibnamefont {Hodgson}}, \bibinfo {author} {\bibfnamefont {J.~D.}\
  \bibnamefont {Ramsden}}, \bibinfo {author} {\bibfnamefont {J.~B.~J.}\
  \bibnamefont {Chapman}}, \bibinfo {author} {\bibfnamefont {P.}~\bibnamefont
  {Lillystone}}, \ and\ \bibinfo {author} {\bibfnamefont {R.~W.}\ \bibnamefont
  {Godby}},\ }\href {\doibase 10.1103/PhysRevB.88.241102} {\bibfield  {journal}
  {\bibinfo  {journal} {Phys. Rev. B}\ }\textbf {\bibinfo {volume} {88}},\
  \bibinfo {pages} {241102} (\bibinfo {year} {2013})}\BibitemShut {NoStop}%
\bibitem [{Note2()}]{Note2}%
  \BibitemOpen
  \bibinfo {note} {We perform calculations for systems of two spinless
  electrons interacting via the appropriately softened Coulomb repulsion \cite
  {SoftenedCoulomb} $u(x,x') = (|x-x'|+1)^{-1}$, and work in Hartree atomic
  units: $m_{\protect \mathrm {e}} = \hbar = e = 4\pi \varepsilon _{0} =
  1$.}\BibitemShut {Stop}%
\bibitem [{Note3()}]{Note3}%
  \BibitemOpen
  \bibinfo {note} {See Supplemental Material at `Link' for the parameters of
  the model systems, and details on our calculations to obtain converged
  results.}\BibitemShut {Stop}%
\bibitem [{\citenamefont {Pines}(1964)}]{Kramers_Kronig}%
  \BibitemOpen
  \bibfield  {author} {\bibinfo {author} {\bibfnamefont {D.}~\bibnamefont
  {Pines}},\ }\href {https://books.google.co.uk/books?id=auNEAAAAIAAJ} {\emph
  {\bibinfo {title} {Elementary excitations in solids: lectures on phonons,
  electrons, and plasmons}}},\ Lecture notes and supplements in physics\
  (\bibinfo  {publisher} {W. A. Benjamin},\ \bibinfo {year} {1964})\BibitemShut
  {NoStop}%
\bibitem [{\citenamefont {Marder}(2010)}]{Kramers_Kronig2}%
  \BibitemOpen
  \bibfield  {author} {\bibinfo {author} {\bibfnamefont {M.}~\bibnamefont
  {Marder}},\ }\href {https://books.google.co.uk/books?id=ijloadAt4BQC} {\emph
  {\bibinfo {title} {Condensed Matter Physics}}}\ (\bibinfo  {publisher}
  {Wiley},\ \bibinfo {year} {2010})\BibitemShut {NoStop}%
\bibitem [{\citenamefont {Ramsden}\ and\ \citenamefont
  {Godby}(2012)}]{iDEA_RE}%
  \BibitemOpen
  \bibfield  {author} {\bibinfo {author} {\bibfnamefont {J.~D.}\ \bibnamefont
  {Ramsden}}\ and\ \bibinfo {author} {\bibfnamefont {R.~W.}\ \bibnamefont
  {Godby}},\ }\href {\doibase 10.1103/PhysRevLett.109.036402} {\bibfield
  {journal} {\bibinfo  {journal} {Phys. Rev. Lett.}\ }\textbf {\bibinfo
  {volume} {109}},\ \bibinfo {pages} {036402} (\bibinfo {year}
  {2012})}\BibitemShut {NoStop}%
\bibitem [{Note4()}]{Note4}%
  \BibitemOpen
  \bibinfo {note} {See Supplemental Material for more details.}\BibitemShut
  {Stop}%
\bibitem [{Note5()}]{Note5}%
  \BibitemOpen
  \bibinfo {note} {For this harmonic well system, at the level of linear
  response theory, only one excitation appears in the absorption
  spectrum.}\BibitemShut {Stop}%
\bibitem [{\citenamefont {Entwistle}\ \emph {et~al.}(2018)\citenamefont
  {Entwistle}, \citenamefont {Casula},\ and\ \citenamefont
  {Godby}}]{Entwistle_LDA2}%
  \BibitemOpen
  \bibfield  {author} {\bibinfo {author} {\bibfnamefont {M.~T.}\ \bibnamefont
  {Entwistle}}, \bibinfo {author} {\bibfnamefont {M.}~\bibnamefont {Casula}}, \
  and\ \bibinfo {author} {\bibfnamefont {R.~W.}\ \bibnamefont {Godby}},\ }\href
  {\doibase 10.1103/PhysRevB.97.235143} {\bibfield  {journal} {\bibinfo
  {journal} {Phys. Rev. B}\ }\textbf {\bibinfo {volume} {97}},\ \bibinfo
  {pages} {235143} (\bibinfo {year} {2018})}\BibitemShut {NoStop}%
\bibitem [{Note6()}]{Note6}%
  \BibitemOpen
  \bibinfo {note} {We define this as a Slater determinant of the occupied KS
  orbitals.}\BibitemShut {Stop}%
\bibitem [{Note7()}]{Note7}%
  \BibitemOpen
  \bibinfo {note} {E. Richardson, private communication}\BibitemShut {NoStop}%
\bibitem [{Note8()}]{Note8}%
  \BibitemOpen
  \bibinfo {note} {As expected for higher energy excited states.}\BibitemShut
  {Stop}%
\bibitem [{Note9()}]{Note9}%
  \BibitemOpen
  \bibinfo {note} {We fold the nonlocal $f_{\protect \mathrm {Hxc}}$ with an
  envelope function that suppresses the more distant nonlocal parts and
  projects the remainder onto the diagonal $x=x'$.}\BibitemShut {Stop}%
\bibitem [{Note10()}]{Note10}%
  \BibitemOpen
  \bibinfo {note} {M. T. Entwistle and R. W. Godby,
  http://dx.doi.org/10.15124/56f576ee-b2de-4ca5-9251-831bfc3cae6f
  (2019).}\BibitemShut {Stop}%
\bibitem [{\citenamefont {Gordon}\ \emph {et~al.}(2005)\citenamefont {Gordon},
  \citenamefont {Santra},\ and\ \citenamefont {K\"artner}}]{SoftenedCoulomb}%
  \BibitemOpen
  \bibfield  {author} {\bibinfo {author} {\bibfnamefont {A.}~\bibnamefont
  {Gordon}}, \bibinfo {author} {\bibfnamefont {R.}~\bibnamefont {Santra}}, \
  and\ \bibinfo {author} {\bibfnamefont {F.~X.}\ \bibnamefont {K\"artner}},\
  }\href {\doibase 10.1103/PhysRevA.72.063411} {\bibfield  {journal} {\bibinfo
  {journal} {Phys. Rev. A}\ }\textbf {\bibinfo {volume} {72}},\ \bibinfo
  {pages} {063411} (\bibinfo {year} {2005})}\BibitemShut {NoStop}%
\end{thebibliography}%
\end{document}